# THE GENERALIZED SIMPLIFIED
# PARTON SHOWER MODEL

R. UGOCCIONI, A. GIOVANNINI, S. LUPIA

*Dipartimento di Fisica Teorica, Università di Torino and*
*INFN, Sezione di Torino, via Giuria 1, 10125 Torino, Italy*

ABSTRACT
We explore the consequences of considering clans real physical objects in the framework of a generalized version of the Simplified Parton Shower model for a single jet. We predict that the average number of clans at fixed energy grows linearly in rapidity and slowly decreases with energy in a fixed rapidity interval.

*Work supported in part by M.U.R.S.T. (Italy) under Grant 1992.*

# THE GENERALIZED SIMPLIFIED PARTON SHOWER MODEL

R. UGOCCIONI*, A. GIOVANNINI and S. LUPIA

*Dipartimento di Fisica Teorica, Università di Torino and*
*INFN, Sezione di Torino, via Giuria 1, 10125 Torino, Italy*

ABSTRACT

We explore the consequences of considering clans real physical objects in the framework of a generalized version of the Simplified Parton Shower model for a single jet. We predict that the average number of clans at fixed energy grows linearly in rapidity and slowly decreases with energy in a fixed rapidity interval.

## 1. The Generalized SPS model

Clan analysis has been a very useful tool for interpreting experimental data in high energy collisions; through its use, two regularities were discovered: first, the average number of clans is approximately constant with energy in a fixed rapidity interval, and second, the average number of clans grows linearly with the rapidity interval at fixed energy. These results have not been explained so far by any theoretical calculation, although they are well reproduced by Monte Carlo simulations. In an attempt to remedy this lack of analytical understanding of experimental facts, we propose to generalize an existing model [1], based on essentials of QCD, by assuming clans as *real* objects. This assumption should be contrasted with the *statistical* clan concept used in the original model.

In the original Simplified Parton Shower (SPS) model [1], we consider an initial parton of maximum allowed virtuality $W$ which splits at virtuality $Q$ into two partons of virtuality $Q_0$ and $Q_1$. We require $Q \geq Q_0 + Q_1$ and $Q_{\min} = 1$ GeV. We define the probability for a parton of virtuality $W$ to split at $Q$, $p(Q|W)$, which is normalized by a Sudakov form factor. The joint probability density $\mathcal{P}(Q_0 Q_1|Q)$ for a parton of virtuality $Q$ to split into two partons of virtuality $Q_0$ and $Q_1$ is defined by $\mathcal{P}(Q_0 Q_1|Q) = p(Q_0|Q)p(Q_1|Q)K(Q)$, where $K(Q)$ is a normalization factor. This general scheme is valid for any splitting function $p(Q|W)$ which is factorizable in terms of its variables $Q$ and $W$. However, for numerical simulations we choose:

$$p(Q|W)dQ = \frac{A}{Q}\frac{(\log Q)^{A-1}}{(\log W)^A}dQ = d\left(\frac{\log Q}{\log W}\right)^A, \tag{1}$$

---

*Work supported in part by M.U.R.S.T. (Italy) under Grant 1992*
* Talk presented by R. Ugoccioni

$A$ being the only free parameter of the model. This form of $p(Q|W)$ was motivated in [1] by our request of simplicity in the structure of the model. Notice that eq. (1) is exactly the virtuality dependence of the standard QCD kernel controlling gluon emission by a parton. For describing the rapidity structure of the model, we use the singular part of the QCD kernel controlling gluon branching:

$$P(z_i)dz_i \propto \left(\frac{1}{z_i} + \frac{1}{1-z_i}\right) dz_i \,. \tag{2}$$

Here $z_i$ is the energy fraction of the $i^{th}$ produced parton ($i = 0, 1$).

We propose now to incorporate in the original SPS model which we have just sketched the concept of clan as follows: we decide to pay attention *for each event* to the ancestor which, splitting $n$ times, gives rise to $n$ subprocesses (one at each splitting) and we identify them with clans. Therefore in this model for a single event the concept of clan is no more as it has been in [2], *i.e.*, only a *statistical* one. In the present picture *the number of clans in each event coincides with the number of splittings of the ancestor, i.e., with the number of steps in the cascade.* It will be shown that this simple consideration is enough to predict that the distribution of clans in full phase space is Poissonian; we stress that this result can be obtained *a priori* in the generalized version of the model, differently from what has been done in [2] where the independent production of clans was an "ansatz" introduced *a posteriori* in order to explain the occurrence of NB regularity.

The generalization of SPS to a genuine Markov process proceeds by assuming that the splitting function of the first step is the same at any step. This fact implies that our kinematical limits are extended from a triangle ($Q_0 + Q_1 \leq Q$) to a square ($Q_0 \leq Q$, $Q_1 \leq Q$), so that subsequent steps are not correlated. We call this new version of the SPS model Generalized Simplified Parton Shower (GSPS) model. It follows that the generating function of the number of clans in full phase space (fps) is given by:

$$g_{\text{clan}}^{\text{fps}}(z, W) = z e^{\lambda_P(\text{fps}, W)(z-1)}, \tag{3}$$

*i.e.*, a shifted Poisson distribution. Accordingly, the average number of clans can easily be calculated by using eq. (1):

$$\bar{N}(\text{fps}, W) = \lambda_P(\text{fps}, W) + 1 \equiv \int_2^W p(Q|Q)dQ + 1 = A \log\left(\frac{\log W}{\log 2}\right) + 1 \,. \tag{4}$$

In order to explain the clan scaling in rapidity, we extend our study to clan distributions in symmetric rapidity intervals $\Delta y = [-y_{\text{cut}}, y_{\text{cut}}]$. We can relate the probability to have $N'$ clans in the symmetric rapidity interval $\Delta y$, $P_{N'}(y_{\text{cut}}, W)$, to the corresponding probability defined in full phase space, $P_N(\text{fps}, W)$, by the relation

$$P_{N'}(y_{\text{cut}}, W) = \sum_{N=N'}^{\infty} \Pi(N', y_{\text{cut}}|N, \text{fps}) P_N(\text{fps}, W), \tag{5}$$

where $\Pi(N', y_{\text{cut}}|N, \text{fps})$ is the conditional probability to have $N'$ clans in $\Delta y$ when one has $N$ clans in full phase space. This new function contains all dynamical information on the *rapidity structure* of the production process. Since *clans are not correlated in full phase space and the only allowed correlations are among particles within the same clan*, the probability $\Pi(N', y_{\text{cut}}|N, \text{fps})$ results to be a positive binomial distribution

$$\Pi(N', y_{\text{cut}}|N, \text{fps}) = \binom{N}{N'} \pi^{N'}(1-\pi)^{N-N'}, \tag{6}$$

where $\pi(y_{\text{cut}}, W)$ is the probability to produce a single clan in the rapidity interval $\Delta y$ from the maximum virtuality $W$. Accordingly, it can be shown that the probability generating function for clans in symmetric rapidity intervals turns out to be

$$g_{\text{clan}}^{\Delta y}(z, W) = [\pi(y_{\text{cut}}, W)z + 1 - \pi(y_{\text{cut}}, W)]e^{\lambda_P(y_{\text{cut}}, W)(z-1)}, \tag{7}$$

with the average number given by $\lambda_P(y_{\text{cut}}, W) = \pi(y_{\text{cut}}, W)\lambda_P(\text{fps}, W)$.

Equation (7), being the sum of two Poissonian distributions (the first a shifted one), has a nice physical meaning: the two terms correspond to the probability of having the ancestor within or outside the given rapidity interval. In addition, it should be noticed that the void probability [3] in a given rapidity interval, $P_0(y_{\text{cut}}, W)$, is different from zero and is given by

$$P_0(y_{\text{cut}}, W) = [1 - \pi(y_{\text{cut}}, W)]e^{-\pi(y_{\text{cut}}, W)\lambda_P(\text{fps}, W)}. \tag{8}$$

This fact is far rich in consequences, which we propose to explore in a forthcoming paper. We also notice that, when $y_{\text{cut}}$ is very small and therefore $\pi(y_{\text{cut}}, W)$ tends to zero, the unshifted Poissonian dominates. Thus, it can be stated that in the smallest rapidity intervals the clan multiplicity is Poissonian to a good approximation. When $y_{\text{cut}}$ tends to full phase space and therefore $\pi(y_{\text{cut}}, W)$ tends to 1, we approach the exact, full phase space, shifted Poisson distribution. We find also that, when $\lambda_P(y_{\text{cut}}, W)$ is sufficiently large, the shifted Poissonian dominates at large $N$ (the tail of the distribution) while the unshifted one dominates at small $N$ (the head of the distribution). This fact might have some consequences in interpreting the anomalies found in NB behavior for small $N$. Furthermore, the average number of clans in a symmetric rapidity interval, $|y| \leq y_{\text{cut}}$, can be calculated from eq. (7) and is given by the following formula:

$$\bar{N}(y_{\text{cut}}, W) = \pi(y_{\text{cut}}, W)\bar{N}(\text{fps}, W). \tag{9}$$

The next step is to calculate the explicit form of $\pi(y_{\text{cut}}, W)$, *i.e.*,

$$\pi(y_{\text{cut}}, W) = \int_{-y_{\text{cut}}}^{y_{\text{cut}}} \tilde{p}(y_{\text{clan}}, W) dy_{\text{clan}}, \tag{10}$$

where $\tilde{p}(y_{\text{clan}}, W)$ is the probability to find a clan of rapidity $y_{\text{clan}}$ generated at any step of the cascade initiated by an ancestor of maximum allowed virtuality $W$. The just mentioned probability can be expressed, within the GSPS, in terms of probability distributions calculable from eq. (1) and (2). It results that the explicit form of $\pi(y_{\text{cut}}, W)$, valid in the central rapidity region, is given by

$$\pi(y_{\text{cut}}, W) = \frac{A}{A-1} \frac{y_{\text{cut}}}{\log W} + \frac{\left[(\log 2W - y_{\text{cut}})^A - (\log 2W + y_{\text{cut}})^A\right]}{2^A (A-1)(\log W)^A}; \tag{11}$$

$$0 < y_{\text{cut}} < \log W - \log 2$$

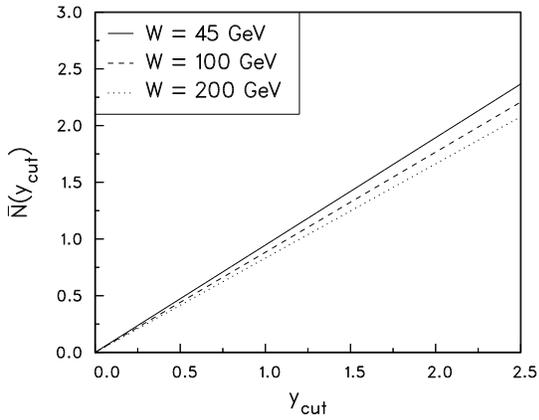 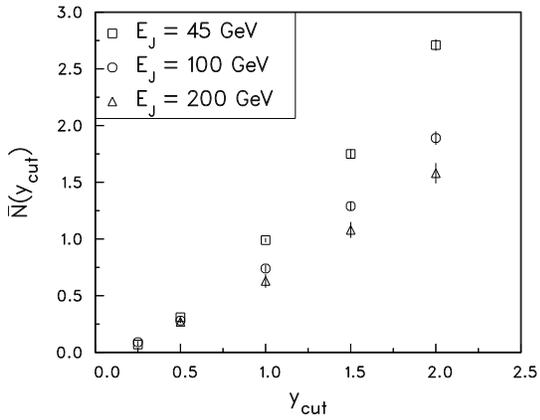

**Figure 1.** Average number of clans $\bar{N}(y_{\text{cut}}, W)$ as a function of rapidity interval $y_{\text{cut}}$ at different ancestor parton energies $W$ in the GSPS model.

**Figure 2.** Average number of clans $\bar{N}(y_{\text{cut}}, W)$ at partonic level as a function of rapidity interval $y_{\text{cut}}$ at different jet energies $E_J$ for lowest energy jets from a sample of 100000 3-jet events generated with JETSET 7.2 Parton Shower at $\sqrt{s} = 1000$ GeV. Jet selection has been performed using the LUCLUS algorithm; partonic level has been obtained via GLPHD.

By inserting eq. (11) and (4) into eq. (9), we determine the behavior of the average number of clans as a function of the rapidity interval and of the energy of the ancestor parton, which is shown for $A = 2$ in Fig. 1. A similar behavior occurs for other values of $A > 0$ and our choice of $A = 2$ should be considered only indicative. As can be seen by inspection, the average number of clans grows linearly with rapidity at fixed energy and is a smoothly decreasing function of energy in a fixed rapidity interval. We point out that no one model to our knowledge has a so sharp *analytical* prediction on clan behavior in single jets. It is clear that our approximation consists in treating produced partons in a jet as pure gluons: our parton shower therefore should be considered a gluon shower. We notice that a similar behavior was seen by us [5] at hadronic level in single lowest energy jets disentangled by using the LUCLUS algorithm from a sample of 3-jet events generated with JETSET 7.2 Parton Shower [6]. We remind that the lowest energy jets in 3-jet events are usually identified with gluon jets. In order to test our predictions in a region where data are not available, we propose to reconstruct the parton level from the above mentioned sample of lowest energy jets in spite of the intrinsic limitations of Monte Carlo models and jet-finding algorithms. In view of NB regularity occurring in this sample [5], we proceed by applying Generalized Local Parton Hadron Duality (GLPHD) prescription [7]. Results at the same jet energies seen in Fig. 1 are shown in Fig. 2: the predicted trend of the GSPS model is confirmed. The above predictions can be tested in the real world at LEP I and LEP II once a definite jet finding algorithm is universally accepted. Were our analytical predictions confirmed in experimental data, this fact would raise the question of the real existence of clans, *i.e.*, groups of particles of common ancestor, in the production process and of their intimate physical nature.

## 2. Acknowledgements

One of us (R.U.) would like to thank prof. M. Block and prof. A. White, and the Local Organizing Committee, for the fruitful and stimulating atmosphere that was created at this symposium.